\documentclass{aastex61}

\newcommand{\mstar}{M_\ast}
\newcommand{\rexp}{r_{\rm exp}}
\newcommand{\stefan}{\sigma_{\rm sb}}

\newcommand{\tevap}{T_{\rm evap}}

\newcommand{\vesc}{v_{\rm esc}}

\received{}
\revised{}
\accepted{\today}
\submitjournal{ApJ}

\shorttitle{Protoplanetary Disk with a Disk Wind}
\shortauthors{Chambers}

\begin{document}

\title{An Analytic Model for an Evolving Protoplanetary Disk with a Disk Wind}

\correspondingauthor{John Chambers}
\email{jchambers@carnegiescience.edu}

\author{John Chambers}
\affil{Carnegie Institution for Science \\
Department of Terrestrial Magnetism, \\
5241 Broad Branch Road, NW, \\
Washington, DC 20015, USA}

%
%
\begin{abstract}
We describe an analytic model for an evolving protoplanetary disk driven by viscosity and a disk wind. The disk is heated by stellar irradiation and energy generated by viscosity. The evolution is controlled by 3 parameters: (i) the inflow velocity towards the central star at a reference distance and temperature, (ii) the fraction of this inflow caused by the disk wind, and (iii) the mass loss rate via the wind relative to the inward flux in the disk. The model gives the disk midplane temperature and surface density as a function of time and distance from the star. It is intended to provide an efficient way to calculate conditions in a protoplanetary disk for use in simulations of planet formation. In the model, disks dominated by viscosity spread radially while losing mass onto the star. Radial spreading is the main factor reducing the surface density in the inner disk. The disk mass remains substantial at late times. Temperatures  in the inner region are high at early times due to strong viscous heating. Disks dominated by a wind undergo much less radial spreading and weaker viscous heating. These disks have a much lower mass at late times than purely viscous disks. When mass loss via a wind is significant, the surface density gradient in the inner disk becomes shallower, and the slope can become positive in extreme cases. 
\end{abstract}

\keywords{accretion, accretion disks; planets and satellites: formation; protoplanetary disks}

%
%
\section{Introduction}
Most stages of planet formation take place in a protoplanetary disk, the cloud of gas and dust that typically surrounds a young star. The density and temperature in a disk vary with distance from the star, and they also vary over time at a single location as the disk evolves and loses mass. These variations affect many aspects of planet formation, including the dynamcs and growth of dust grains, the chemical composition of solids at a given location, the efficiency of planetesimal formation and growth, and the migration of planetary orbits \citep{birnstiel:2012, dodson-robinson:2009, ormel:2013, ogihara:2018}.

A good model for a protoplanetary disk is clearly important for understanding planet formation. Studies of planetary growth have used a variety of disk models of varying complexity \citep{ida:2004, mcneil:2005, mordasini:2009, chambers:2016}. The simplest case is a disk with fixed surface density and temperature profiles, usually taken to be power laws. Slightly more realistic models use surface density profiles that decrease over time, including the self-similar solutions for a viscously evolving disk developed by \citet{lynden-bell:1974}. 

More complicated disk models calculate the temperature and surface density evolution self-consistently taking into account heating by viscous accretion and irradiation by the central star \citep{bell:1997, bitsch:2013}. However, such models have two potential drawbacks. Firstly, they can be computationally expensive, which may make them impractical for studies that consider a large number of planet-formation simulations. Secondly, complicated models may introduce unrealistic behavior that is hard to identify. This is a particular worry since the main physical process driving disk evolution is uncertain.

The computational cost of disk models is particularly relevant to population synthesis studies of planet formation. These studies use a relatively simple model for planetary growth and orbital evolution that can be run thousands or millions of times \citep{mordasini:2009, chambers:2018}. The resulting planetary systems can then be compared with the Solar System or the observed distribution of extrasolar planets in order to constrain unknown aspects of the model.

An ideal protoplanetary disk model for a population synthesis study is one that is analytic, contains relatively few free parameters, and captures the features of real disks that are most relevant to planet formation. \citet{stepinski:1998} developed such a model that describes the temperature and surface density in a viscously evolving disk heated by the energy generated by the viscosity. The model uses the mass loss rate onto the star and conservation of angular momentum to determine the time evolution of the disk. \citet{chambers:2009} extended this to a disk heated by viscosity and radiation from the star.

Both these studies assume that the only factors driving disk evolution are viscosity and mass falling onto the star. The viscosity is calculated using the popular alpha-disk model in which viscosity is assumed to be proportional to the sound speed and scale height of the gas \citep{shakura:1973}. The source of visosity is not specified, but could be due to magnetorotational instability or hydrodynamic instabilties for example \citep{balbus:1991, nelson:2013}.

Recently, it has been suggested that viscosity is not the only process driving disk evolution. The observed properties of resolved protoplanetary disks do not appear to follow the correlations expected for purely viscous disks, suggesting that an additional process is operating \citep{rafikov:2017}. The small scale height of dust grains in HL Tau implies turbulence is weak in this disk. If viscosity is associated with turbulence, the implied viscosity is too weak to drive the observed mass accretion rate onto the star \citep{pinte:2016}. There is some additional support for low turbulence levels from the observed line widths of gas molecules in disks \citep{flaherty:2015}.

A promising disk-evolution mechanism is a disk wind driven by interactions with a magnetic field \citep{suzuki:2010, bai:2013, simon:2013}. A disk with a strong wind behaves differently than a purely viscous disk for several reasons. The wind ejects material from the disk, providing a second mass sink in addition to accretion onto the star. The wind also exerts a torque on the remaining material. This means that the disk's angular momentum is not conserved, in contrast to cases where the evolution is controlled by viscosity alone. The wind torque causes gas in the disk to flow inwards in addition to any motion driven by viscosity. Disks with a significant wind are also likely to be cooler than comparable viscous disks because the inflow driven by the wind does not lead to viscous heating \citep{suzuki:2016}.

The possibility that disk winds are important means we should consider winds in disk models that are used in studies of planet formation. In this paper, we develop an analytic model for an evolving disk subject to a disk wind and viscosity, assuming that both are driven by interactions with a magnetic field. The model is primarily intended for use with models of planet formation, but it could also be used as a starting point for studies of disks themselves.

The rest of this paper is organized as follows. Section~2 contains a detailed derivation of the model, and the main equations are summarized in Section~3. In Section~4, we look at some example disk evolutions, highlighting the main differences between purely viscous disks and those with a wind. Section~5 contains a summary.

%
%
\section{Model Derivation}
In this section, we derive a model for an evolving protoplanetary disk with a disk wind. Those readers who are not interested in the details of the derivation can skip to the next section where the model is summarized.

The analytic disk evolution models of \citet{stepinski:1998} and \citet{chambers:2009} calculated a series of steady-state solutions for the disk surface density and temperature profiles, and linked them together to get the time evolution using conservation of angular momentum. When a disk wind is present, angular momentum is no longer conserved. Here, we use a different approach and look for an approximate solution to the time-dependent equation describing the surface density evolution due to viscosity and a disk wind. Some short-cuts are necessary to make this equation analytically tractable, but we believe these are justified given the current large uncertainties in the physics of disk evolution.

Following \citet{suzuki:2016}, we assume that the disk wind and viscosity are driven by interactions with a magnetic field, and that the evolution of the surface density $\Sigma$ can be expressed as
\begin{equation}
\frac{\partial\Sigma}{\partial t}=
\frac{2}{r}\frac{\partial}{\partial r}\left[
\frac{1}{r\Omega}\frac{\partial}{\partial r}
\left(\bar\alpha_{r\phi}r^2\Sigma c_s^2\right)\right]
+\frac{2}{r}\frac{\partial}{\partial r}\left[
\frac{\bar\alpha_{\phi z}r\rho c_s^2}{\Omega}\right]
+C_W\rho c_s
\end{equation}
where $r$ is the distance from the star, $\Omega\propto r^{-3/2}$ is the Keplerian angular velocity, $c_s\propto T^{1/2}$ is the sound speed, $T$ is the temperature, and $\rho\simeq\Sigma\Omega/(2c_s)$ is the gas density. The three quantities $\bar\alpha_{r\phi}$, $\bar\alpha_{\phi z}$ and $C_W$ are parameters that depend on the magnetic field.

We rewrite this equation as
\begin{equation}
\frac{\partial\Sigma}{\partial t}=
\frac{3}{r}\frac{\partial}{\partial r}\left[
r^{1/2}\frac{\partial}{\partial r}
\left(r^{1/2}\nu \Sigma\right)\right]
+\frac{1}{r}\frac{\partial}{\partial r}\left(r v_w\Sigma
\right)-\dot\Sigma_w
\label{eq_main}
\end{equation}
where $\nu$ is the viscosity, $v_W$ is the inward radial velocity induced by the disk wind, and $\dot\Sigma_w$ is the rate of change of the surface density due to mass ejected by the disk wind.

We normalize the distance using a reference radius $r_0$ which we set to 1 AU unless noted otherwise. The initial surface density at $r_0$ is $\Sigma_0$. We also normalize the temperature to $T_0$ which is the temperature the disk would have at $r_0$ due solely to stellar irradiation. The total inward velocity of disk material at $r_0$ and $T_0$ is $v_0$, and the fraction of this velocity (at $r_0$ and $T_0$) due to the disk wind is $f_w$. 

Using these normalizations, we can express $\nu$ and $v_w$ and $\dot\Sigma_w$ as follows:
\begin{eqnarray}
\nu&=&\frac{2}{3}(1-f_w)r_0v_0\left(\frac{r}{r_0}\right)^{3/2}\left(\frac{T}{T_0}\right)
\nonumber \\
v_w&=&f_wv_0\left(\frac{T}{T_0}\right)^{1/2} \nonumber \\
\dot\Sigma_w&=&\frac{Kf_wv_0\Sigma}{r_0} \left(\frac{r}{r_0}\right)^{-3/2}
\end{eqnarray}
where $K$ is a constant, and conservation of angular momentum implies that
\begin{equation}
K=\frac{1}{2}\left(\frac{r}{r_0}\right)^{-1/4}
\left(\frac{\Omega r}{\vesc-\Omega r}\right)
\end{equation}
where $\vesc$ is the tangential velocity of the escaping disk wind.

The model parameters $v_0$, $f_W$ and $K$ are related to the 3 parameters $C_{w,0}$, $\bar\alpha_{r\phi}$ and $\bar\alpha_{\phi z}$ used by \citet{suzuki:2016} as follows
\begin{eqnarray}
\bar\alpha_{r\phi}&=&\frac{(1-f_w)r_0v_0\Omega_0}{c_{s0}^2}
\nonumber \\
\bar\alpha_{\phi z}&=&\frac{f_w v_0}{c_{s0}}
\nonumber \\
C_W&=&\frac{2Kf_w v_0}{r_0\Omega_0}
\end{eqnarray}
where $\Omega_0$ and $c_{s0}$ are the values of $\Omega$ and $c_s$ at the reference radius and temperature.

Following \citet{lecar:2006}, the temperature at the midplane of the disk is
\begin{equation}
T^4=T_0^4\left(\frac{r}{r_0}\right)^{-2}
+\frac{3G\mstar F}{8\pi\stefan r^3}\times\frac{3\kappa\Sigma}{8}
\end{equation}
where $\mstar$ is the mass of the star, $\stefan$ is the Stefan-Boltzmann constant, $\kappa$ is the opacity of the disk, and $F\simeq3\pi\Sigma\nu$ is the mass flux due to viscosity. The first term on the righthand side of this equation is the temperature due to stellar irradiation (assumed to be proportional to $r^{-1/2}$), and the second term is due to viscous heating.

Using normalized variables, and the expression for $\nu$, we can rewrite the temperature equation as
\begin{equation}
\theta^4=x^{-4}+\sigma^2\theta x^{-3}\left(\frac{\kappa}{\kappa_0}\right)
\label{eq_temperature}
\end{equation}
where
\begin{eqnarray}
x&=&\left(\frac{r}{r_0}\right)^{1/2} \nonumber \\
\sigma&=&\frac{A\Sigma}{\Sigma_0} \nonumber \\
\theta&=&\frac{T}{T_0}
\end{eqnarray}
and
\begin{equation}
A^2=\frac{9(1-f_w)G\mstar\kappa_0\Sigma_0^2v_0}{32\sigma_{sb} r_0^2T_0^4}
\end{equation}
where $\kappa_0$ is a reference value of the opacity, discussed below. We note that  when $\sigma<1$, the midplane temperature is mainly determined by stellar irradiation. Viscosity is the dominant source of heating when $\sigma>1$.

Substituting these expressions into Eqn.~\ref{eq_main}, the surface density evolution is
\begin{equation}
Lx^3 \frac{\partial\sigma}{\partial t}=
\frac{\partial^2}{\partial x^2} \left(x^4\sigma\theta\right)
+J\frac{\partial}{\partial x} \left(x^2\sigma \theta^{1/2}\right)
-2JK\sigma
\label{eq_true}
\end{equation}
where
\begin{eqnarray}
L&=&\frac{2r_0}{v_0(1-f_w)} \nonumber \\
J&=&\frac{f_w}{(1-f_w)}
\end{eqnarray}

The evolution of the disk depends on 3 parameters: $f_w$, $K$ and $v_0$ (or equivalently $J$, $K$ and $L$). We assume that these parameters are constant, although they may vary with time and location in real disks. 

To make the evolution equation easier to solve analytically, we modify the exponents of $x$ and $\theta$ somewhat to give
\begin{equation}
Lx^{7/2}\frac{\partial}{\partial t}(\sigma\theta^{1/2})=
\frac{\partial^2}{\partial x^2} \left(x^3\sigma\theta^{1/2}\right)
+J\frac{\partial}{\partial x} \left(x^2\sigma \theta^{1/2}\right)
-2JKx\sigma\theta^{1/2}
\label{eq_simplified}
\end{equation}
Note that the $x$ dependence of each term has been altered in the same sense as the temperature $\theta$, which should partially compensate for these changes. Given other large uncertainties in the model, we believe these changes are justified and the modified equation should retain the main features of the disk evolution. We investigate the effect of these alterations in the Appendix.

We change variables using
\begin{equation}
p=x^3\sigma\theta^{1/2}
\label{eq_pdef}
\end{equation}
which gives  
\begin{equation}
Lx^{1/2}\frac{\partial p}{\partial t}=\frac{d^2p}{dx^2}+Jx^{-1}\frac{dp}{dx}-J(1+2K)x^{-2}p
\end{equation}

We look for a self-similar solution of the form $p=\Theta(t)Y(y)$ where $y=x/s(t)$, which gives
\begin{equation}
Ly^{1/2}Ys^{5/2}\Theta^{-1}\frac{d\Theta}{dt}
-Ls^{3/2}y^{3/2}\frac{ds}{dt}\frac{dY}{dy}
=
\frac{d^2Y}{dy^2}
+Jy^{-1}\frac{dY}{dy}
-J(1+2K)y^{-2}Y
\end{equation}

We set $B=Ls^{5/2}\Theta^{-1}(d\Theta/dt)$ and $C=Ls^{3/2}(ds/dt)$ where $B$ and $C$ are constants, so that
\begin{equation}
\frac{d^2Y}{dy^2}
+Jy^{-1}\frac{dY}{dy}
+Cy^{3/2}\frac{dY}{dy}
-J(1+2K)y^{-2}Y
-By^{1/2}Y
=0
\end{equation}

This equation has the following analytic solution
\begin{equation}
Y=y^b\exp(-y^c)
\end{equation}
where
\begin{eqnarray}
b&=&\frac{(1-J)+[(1+J)^2+8JK]^{1/2}}{2}
\nonumber \\
c&=&\frac{5}{2} \nonumber \\
B&=&c(1-b-c-J) \nonumber \\
C&=&\frac{5}{2}
\end{eqnarray}

Given $B$ and $C$, we can solve for $s$ and $\Theta$ to give
\begin{eqnarray}
s&=&s_0\left(1+\frac{t}{\tau}\right)^{2/5}
\nonumber \\
\Theta&=&\Theta_0
\left(1+\frac{t}{\tau}\right)^{4B/25}
\end{eqnarray}
where
\begin{equation}
\tau=\frac{4Ls_0^{5/2}}{25}
=\frac{8r_0s_0^{5/2}}{25v_0(1-f_w)}
\end{equation}

To obtain the disk temperature, we assume an opacity law of the form
\begin{equation}
\kappa=\frac{\kappa_0}{1+V^{-n}\theta^n}
\end{equation}
where $n$ is a large positive number \citep{stepinski:1998}, and
\begin{equation}
V=\frac{\tevap}{T_0}
\end{equation}
where $\tevap=1500$ K is the dust evaporation temperature. Thus, we assume the opacity is roughly constant below the dust evaporation temperature, and rapidly falls to zero at $T\ge \tevap$. More complicated opacity laws are possible for $T<\tevap$, but we believe this complexity is not justified due to large uncertainties in the degree of aggregation and radial redistribution of dust in real disks.

From Eqn.~\ref{eq_temperature}, the temperature can be expressed as
\begin{equation}
\theta^4 x^4=1+
\frac{\sigma^2 \theta x}{(1+V^{-n}\theta^n)}
\end{equation}

We find that the following is a good approximation (noting that $Vx>1$ in almost all cases):
\begin{equation}
\theta^3\simeq V^3\left(\frac{1+\sigma^2}{\sigma^2+V^3x^3}\right)
\label{eq_theta}
\end{equation}
which forces the temperature to asymptote to $\tevap$ close to the star.

We also need to find the normalized surface density $\sigma$ given $p$. From Eqn.~\ref{eq_pdef}, we have
\begin{equation}
p=x^3\sigma\theta^{1/2}
\simeq V^{1/2}x^3\sigma
\left(\frac{1+\sigma^2}{\sigma^2+V^3x^3}\right)^{1/6}
\end{equation}

We find that the following approximate inversion for $\sigma$ works well
\begin{equation}
\sigma\simeq q\left[
\frac{1+V^{-2}x^{-2}q}{1+q}
\right]^{1/4}
\label{eq_sigma}
\end{equation}
where $q=px^{-5/2}$

The remaining constants $s_0$ and $\Theta_0$ can be obtained from the initial conditions. If we choose an initial exponential turnover radius for the disk $\rexp$, then $s_0$ is given by $\rexp=r_0s_0^2$. In addition, $\Theta_0$ is given by
\begin{equation}
\Theta_0=p_0s_0^b\exp(s_0^{-5/2})
\end{equation}
where
\begin{equation}
p_0=AV^{1/2}\left(\frac{1+A^2}{A^2+V^3}\right)^{1/6}
\end{equation}

For the parameters we consider in this paper, most of the disk mass is contained in the outer region where the temperature is mainly determined by stellar irradiation. We can get a rough estimate for the total disk mass by assuming the entire disk is radiative. In this case, we have
\begin{equation}
\sigma\simeq px^{-5/2}
\end{equation}
so that
\begin{equation}
M=\int_0^\infty 2\pi r \Sigma dr
\simeq M_0
\left(1+\frac{t}{\tau}\right)^{-m}
\end{equation}
where
\begin{equation}
m=2(b+J)/5
\end{equation}
and
\begin{equation}
M_0=
\frac{4\pi r_0^2\Sigma_0s_0^{b+3/2}p_0 I_b}{A}
\exp(s_0^{-5/2})
\end{equation}
and
\begin{eqnarray}
I_b&=&\int_0^\infty y^{b+1/2} \exp(-y^{5/2}) dy
\nonumber \\
&\simeq& 0.38
\hspace{40mm} 1\le b\le 3
\end{eqnarray}
where these $b$ values span the likely range for real disks.

%
%
\section{Model Summary}
In the previous section, we developed an analytic model for an evolving protoplanetary disk subject to viscous forces and a disk wind. Here we summarize the main formulae for the model.

We consider a disk with initial surface density $\Sigma_0$ at a reference radius $r_0$. The initial temperature at $r_0$ due solely to stellar irradiation would be $T_0$, although viscous heating will typically raise the temperature above this. Material flows inward in the inner disk. At $r_0$ and $T_0$ the inward radial velocity is $v_0$, and the fraction of this inward velocity caused by the disk wind is $f_w$. The surface density decreases everywhere as mass escapes in the disk wind, where the tangential wind velocity is $\vesc(r)$. The mass loss rate is characterized by $K$, such that
\begin{equation}
\left(\frac{\partial\Sigma}{\partial t}\right)_{\rm wind}=-\frac{Kf_wv_0\Sigma}{r_0}\left(\frac{r}{r_0}\right)^{-3/2}
\end{equation}
where
\begin{equation}
K=\frac{1}{2}\left(\frac{r}{r_0}\right)^{-1/4}
\left(\frac{\Omega r}{\vesc-\Omega r}\right)
\end{equation}
where $\Omega$ is the Keplerian orbital frequency. The evolution is controlled by three parameters, $v_0$, $f_w$ and $K$, which are assumed to be constant. 

The surface density and temperature at time $t$ and radial distance $r$ from the star are given by
\begin{eqnarray}
\Sigma&=&\frac{\Sigma_0}{A}px^{-5/2}\left[
\frac{1+V^{-2}px^{-9/2}}{1+px^{-5/2}}
\right]^{1/4}
\nonumber \\
T&=&\tevap\left(\frac{\sigma^2+1}{\sigma^2+V^3x^3}\right)^{1/3}
\label{eq_sigma_result}
\end{eqnarray}
where
\begin{eqnarray}
x&=&\left(\frac{r}{r_0}\right)^{1/2} \nonumber \\
\sigma&=&\frac{A\Sigma}{\Sigma_0} \nonumber \\
V&=&\frac{\tevap}{T_0} \nonumber \\
A^2&=&\frac{9(1-f_w)G\mstar\kappa_0\Sigma_0^2v_0}{32\sigma_{sb} r_0^2T_0^4}
\end{eqnarray}
where $\mstar$ is the stellar mass, $\stefan$ is the Stefan-Boltzmann constant, $\tevap$ is the dust evaporation temperature, $\kappa_0$ is the (constant) opacity when $T<\tevap$, and
\begin{equation}
p=p_0\left(1+\frac{t}{\tau}\right)^nx^b
\exp\left[\left(\frac{1}{s_0}\right)^{5/2}
-\left(\frac{x}{s_0}\right)^{5/2}
\left(1+\frac{t}{\tau}\right)^{-1}\right]
\end{equation}
where
\begin{eqnarray}
p_0&=&AV^{1/2}\left(\frac{A^2+1}{A^2+V^3}\right)^{1/6}
\nonumber \\
n&=&-1-\frac{2}{5}[(1+J)^2+8JK]^{1/2}
\nonumber \\
b&=&\frac{(1-J)}{2}+\frac{1}{2}[(1+J)^2+8JK]^{1/2}
\nonumber \\
J&=&\frac{f_w}{(1-f_w)}
\nonumber \\
\tau&=&\frac{8r_0s_0^{5/2}}{25v_0(1-f_w)}
\end{eqnarray}

The surface density in the outer disk decays exponentially with distance. The initial exponential turnover distance is $\rexp=r_0s_0^2$, which determines the value of $s_0$. The initial disk mass $M_0$ is approximately related to $\Sigma_0$ in Eqn.~\ref{eq_sigma_result}
by
\begin{equation}
M_0\simeq
\frac{4\pi r_0^2\Sigma_0s_0^{b+3/2}p_0 I_b}{A}
\exp(s_0^{-5/2})
\end{equation}
where $I_b$ is an integral that depends very weakly on $b$, which we take to be a constant:
\begin{equation}
I_b\simeq 0.38
\end{equation}

The disk mass at time $t$ is approximately given by
\begin{equation}
M\simeq M_0
\left(1+\frac{t}{\tau}\right)^{-m}
\label{eq_mrough}
\end{equation}
where
\begin{equation}
m=2(b+J)/5
\end{equation}

\begin{figure}
\plotone{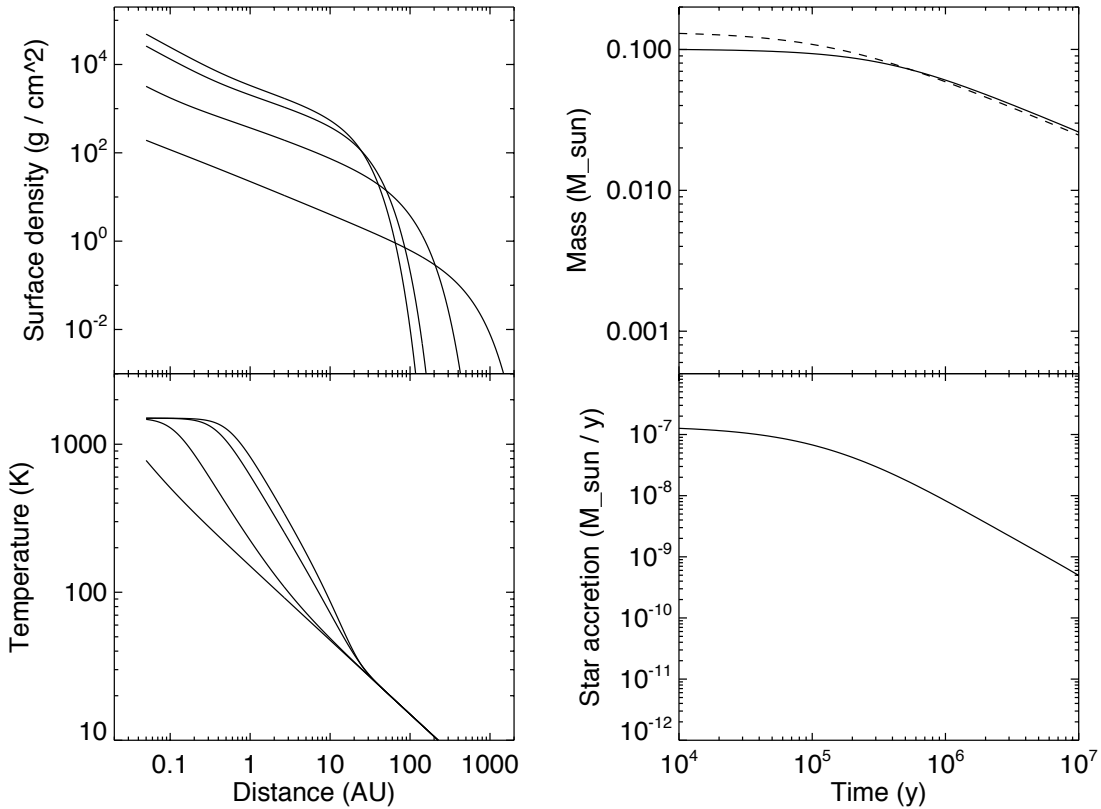}
\caption{Evolution of a disk subject to viscous forces only. The initial mass and exponential radius are 0.1 solar masses and 15 AU, respectively, and $v_0=30$ cm/s. The upper and lower lefthand panels show the surface density and temperature at 0.01 My, 0.1 My, 1 My and 10 My, with lower curves corresponding to later times. The upper right panel shows the total disk mass (solid curve), and a rough estimate for the disk mass assuming it is entirely radiative (dashed curve). The lower right panel shows the mass accretion rate onto the star.}
\end{figure}

%
%
\section{Examples}
In this section, we look at several examples of the disk model described above. We will focus on the effects of the two disk-wind parameters, $f_w$ and $K$, that describe the relative importance of the disk wind compared to viscous forces, and the degree to which the wind erodes mass from the disk, respectively.

In the following four examples, we consider a disk with an initial mass of 0.1 solar masses, and an initial exponential radius $\rexp=15$ AU.  (This value of $\rexp$ is chosen to emphasize the different effects of viscosity and the disk wind in the examples below, although many real disks may begin with larger radii.) The central star has a mass of 1 solar mass. For the purpose of calculating the mass accretion rate onto the star, we assume that the inner edge of the disk is at 0.05 AU. The temperature at 1 AU due to stellar irradiation alone is 150 K \citep{chiang:1997}, and this remains constant over time. The dust evaporation temperature is $\tevap=1500$ K. At temperatures below $\tevap$, the opacity of the disk is $\kappa\simeq\kappa_0=0.1$ cm$^2$/g. 

We adopt a value for $\kappa_0$ that is much lower than that due to interstellar dust (typically a few cm$^2$/g for a gas-to-dust ratio of 100) for two reasons. Firstly, we assume that a substantial amount of dust aggregation has taken place, so the total surface area of the grains is a lot less than for interstellar dust \citep{ormel:2013}.  Secondly, we assume that heating due to viscous accretion is concentrated some distance from the disk midplane (at a few scale heights). This means the heat can escape vertically more easily, and so the midplane temperature is less than it would be if the heat were released at the midplane \citep{mori:2019}.

\begin{figure}
\plotone{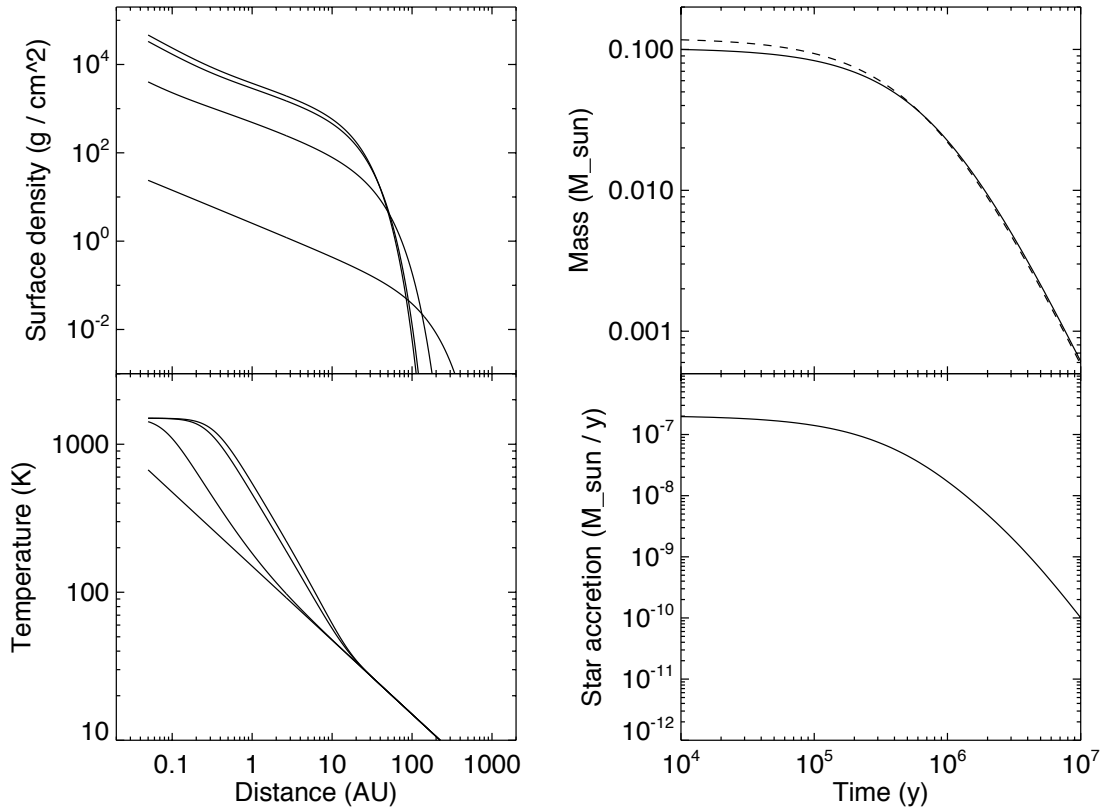}
\caption{Evolution of a disk dominated by a fast disk wind, with negligible mass loss due to the wind ($v_0=30$ cm/s and $f_w=0.8$). The initial mass and exponential radius are 0.1 solar masses and 15 AU, respectively. The upper and lower lefthand panels show the surface density and temperature at 0.01 My, 0.1 My, 1 My and 10 My, with lower curves corresponding to later times. The upper right panel shows the total disk mass (solid curve), and a rough estimate for the disk mass assuming it is entirely radiative (dashed curve). The lower right panel shows the mass accretion rate onto the star.}
\end{figure}

%
%
\subsection{Purely Viscous Disk}
Figure 1 shows the evolution of a purely viscous disk with no disk wind. The inflow velocity at $r_0=1$ AU and $T_0=150$ K is $v_0=30$ cm/s, and the initial surface density at 1 AU is $\Sigma_0=3450$ g/cm$^2$. (This corresponds to an initial stellar mass accretion rate of $1.5\times 10^{-7}M_\odot$/y, and $\alpha\simeq0.01$ for an alpha disk, with $\alpha$ independent of radius.) The upper, left panel of the figure shows the surface density profile at four times: $10^4$, $10^5$, $10^6$ and $10^7$ years, with lower values of $\Sigma$ in the inner disk corresponding to later times. The surface density in the inner parts of the disk declines monotonically over time as the disk spreads radially and mass flows onto the star. The surface density inside 10 AU falls by roughly two orders of magnitude over 10 million years.

In the inner disk, the mass flux is roughly independent of radial distance $r$. The surface density $\Sigma$ profile has several different slopes depending on the temperature. In the innermost disk, at early times, the temperature is close to the dust evaporation temperature, and $\Sigma\propto r^{-1}$. At somewhat larger distances, where temperatures are mainly controlled by viscous heating, the profile is shallower such that $\Sigma\propto r^{-9/16}$. Further out, the temperature is mainly set by stellar irradiation and the surface density profile steepens slightly, with $\Sigma\propto r^{-3/4}$. In the outer disk, where the disk is spreading, the surface density profile steepens with distance, and $\Sigma\propto \exp(-r^{5/4})$. 

The different surface-density profile regimes are reflected in differences in the temperature profile slopes, shown in the lower, left panel of Figure~1. This is because the inward mass flux is almost independent of distance, so $T$ and $\Sigma$ are correlated. In the outer, radiation-dominated region, $T\propto r^{-1/2}$, while the profile steepens in the viscously heated region so that $T\propto r^{-7/8}$. In the innermost region, where temperatures approach the dust evaporation temperature, $T\rightarrow\tevap$ as changes in the viscous heating rate are offset by changes in the dust opacity.

The surface density profiles described above can be obtained directly from Eqn.~\ref{eq_sigma}, noting that $q\propto r^{-3/4}$ (since $b=1$ and $p\propto r^{1/2}$) in the absence of mass loss via a disk wind (i.e. when $K=0$). Situations dominated by viscous heating and irradiation corrrespond to $q>1$ and $q<1$ respectively. The temperature profiles can then be found using Eqn.~\ref{eq_theta} using the fact that $\sigma>1$ for viscously heated regions, and $\sigma<1$ for irradiation-dominated regions.

The upper, right panel of Figure~1 shows the total disk mass. The solid curve shows an accurate measure calculated by sampling the model surface density at 2000 radial locations. The dashed line shows the approximate estimate of the mass using Eqn.~\ref{eq_mrough}, determined by assuming the entire disk is radiatively heated. The rough estimate is somewhat too high at early times, but accurate from 0.3 My onwards. It is notable that the disk mass declines by only about a factor of 4 over $10^7$ years. This contrasts with the behavior of the surface density in the inner disk, which declines by two orders of magnitude. This shows that the surface density decrease is mainly caused by viscous spreading of the disk, while mass loss onto the star is a secondary effect.

The lower, right panel of Figure~1 shows the mass accretion rate onto the star. There is no mass loss via a disk wind, so this is also a good measure of the mass flux throughout the inner part of the disk. The initial mass accretion rate is about $1.5\times 10^{-7}M_\odot$/y. By 10 My, the mass accretion rate has declined to about $5\times 10^{-10} M_\odot$/y, even though the disk still contains about $0.026$ solar masses of material---almost 30 times Jupiter's mass. This demonstrates the importance of viscous spreading as the main factor driving the disk's evolution in this case.

Overall, the evolution of the disk is similar to the behavior of the self-similar solution for a disk with a fixed temperature profile described by \citet{lynden-bell:1974}. The main difference in the surface density behavior is that the profile shown in Figure~1 is somewhat shallower. For a temperature profile $T\propto r^{-1/2}$, the self-similar model of \citet{lynden-bell:1974} predicts $\Sigma\propto r^{-1}$, compared to $\Sigma\propto r^{-3/4}$ in the radiatively heated region in the model described here. This difference is attributable to the approximations made in Eqn.~\ref{eq_simplified} in order to make the problem analytically tractable in the general case where a wind is also present.

%
%
\subsection{Wind Dominated Disk with a Fast Wind}
We now consider a case dominated by a disk wind. Figure~2 shows the evolution of a disk with a fast disk wind that carries away a large amount of angular momentum but negligible mass.
Here, we use the term ``fast wind'' to denote a situation in which we assume that only a small amount of mass is ejected per unit angular momentum removed by the wind. Later, we will consider a ``slow wind'' where a large amount of mass is removed per unit angular momentum removed by the wind.

The initial mass and exponential radius of the disk are 0.1 solar masses and 15 AU, which are the same as the case shown in Figure~1. (The initial surface density at 1 AU is slightly different: $\Sigma_0=3830$ g/cm$^2$.) The inflow velocity at $r_0=1$ AU and $T_0=150$ K is also the same as before: $v_0=30$ cm/s. In this case, 80\% of the inflow at $r_0$ and $T_0$ is driven by the disk wind rather than viscosity. Thus $f_w=0.8$. (The viscosity is equivalent to $\alpha\simeq0.002$ for an alpha disk.) We set $K=0$ since we assume mass loss due to the wind is negligible.

The upper, left panel of Figure 2 shows the surface density profile at $10^4$, $10^5$, $10^6$ and $10^7$ years. The surface density $\Sigma$ in the inner disk declines monotonically with time, and the disk spreads radially outwards, as before. The slopes of $\Sigma$ are similar in the two cases. (This is true for any case with $K=0$). The rate of decrease in the surface density in the inner disk is also similar in the two cases for the first million years. 

However, there are clear differences from the case shown in Figure~1. The degree of radial spreading is much less in Figure~2 than in Figure~1. This is because the viscosity in the outer disk is roughly 5 times smaller. The wind-driven disk has a lower surface density at late times. By $10^7$ years, the surface density inside 10 AU has fallen by roughly 3 orders of magnitude compared to 2 orders of magnitude in the purely viscous disk. This different behavior is also apparent in the disk mass, shown in the upper, right panel of Figure~2. After 10 million years, the disk mass has fallen to $6.1\times 10^{-4}$ solar masses compared with $0.026$ solar masses in Figure~1. 

These differences arise mainly due to the reduced amount of radial spreading when a disk wind drives the evolution. Weaker spreading means that more mass remains in the inner disk where it can accrete onto the star. This allows high rates of mass accretion to continue for longer. For example, the mass loss rate onto the star at 1 My is twice as high in Figure~2 (lower, righthand panel) as in Figure~1. The flip side of this early, efficient accretion is that both the disk mass and mass accretion rate are lower at later times. 

\begin{figure}
\plotone{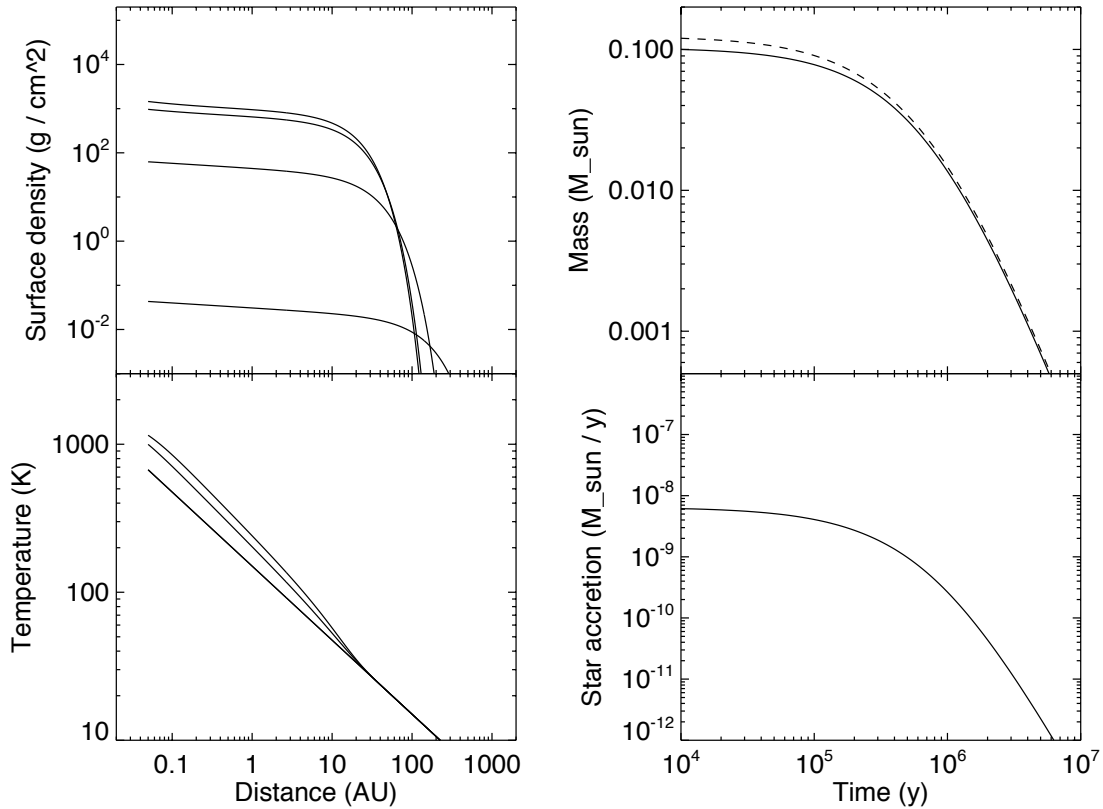}
\caption{Evolution of a disk dominated by a slow disk wind with substantial mass loss via the wind ($v_0=30$ cm/s, $f_w=0.8$, and $K=1$). The initial mass and exponential radius are 0.1 solar masses and 15 AU, respectively. The upper and lower lefthand panels show the surface density and temperature at 0.01 My, 0.1 My, 1 My and 10 My, with lower curves corresponding to later times. The upper right panel shows the total disk mass (solid curve), and a rough estimate for the disk mass assuming it is entirely radiative (dashed curve). The lower right panel shows the mass accretion rate onto the star.}
\end{figure}

The lower, left panel of Figure~2 shows the temperature evolution. This is qualitatively similar to Figure~1. The disk is divided into three regions as before: (i) an outer region dominated by stellar irradiation; (ii) an intermediate region with a steeper temperature profile where heating is mainly caused by viscosity, and (iii) an innermost region where dust is evaporating, which only appears at early times. There are some differences. Temperatures in the inner disk are generally lower in Figure~2 than Figure~1 due to the fact that most of the inflow is caused by a wind which doesn't heat the disk. This difference is partially offset by the fact that mass accretion rates are higher at early times in Figure~2.

The much greater mass depletion at late times in Figure 2 compared to Figure~1 has implications for disk dispersal. The purely viscous disk shown in Figure~1 still contains 0.026 solar masses at 10 My, which is longer than the lifetime of most disks \citep{haisch:2001}. This suggests an additional mechanism, such as photoevaporation, is needed to remove mass late in the lifetime of a disk \citep{gorti:2009}. In contrast, the disk with a wind, shown in Figure~2 has a much lower mass at 10 My. Additional mass loss mechanisms may be unimportant when a strong wind is present.

%
%
\subsection{Wind Dominated Disk with a Slow Wind}
We now consider a case with a slow disk wind where the wind removes a large amount of mass as well as angular momentum. The initial disk mass, exponential radius and inflow velocity $v_0$ are the same as the previous cases. We also set $f_w=0.8$, the same as in Figure~2, so that 80\% of the inflow (at $r_0$ and $T_0$) is driven by the wind. (The viscosity is equivalent to $\alpha\simeq0.002$ for an alpha disk, as in the previous case.) However, we make $K=1$, which means that a large portion of the inflowing mass is ejected by the wind before reaching the star. In order to keep the initial mass at 0.1 solar masses, this requires setting $\Sigma_0=1040$ g/cm$^2$.

Figure~3 shows the evolution in this case. It is immediately apparent that the surface density profile is very different than the previous cases. The profile is almost flat across the entire region interior to 10 AU at all stages of the evolution. In the region dominated by stellar irradiation $\Sigma\propto r^{-0.11}$, while in the viscously heated region $\Sigma\propto r^{-0.14}$. The surface density across much of this region is also much lower than before. For example, the surface density at 1~AU at 1~My is an order of magnitude smaller in Figure~3 than Figure~2.

\begin{figure}
\plotone{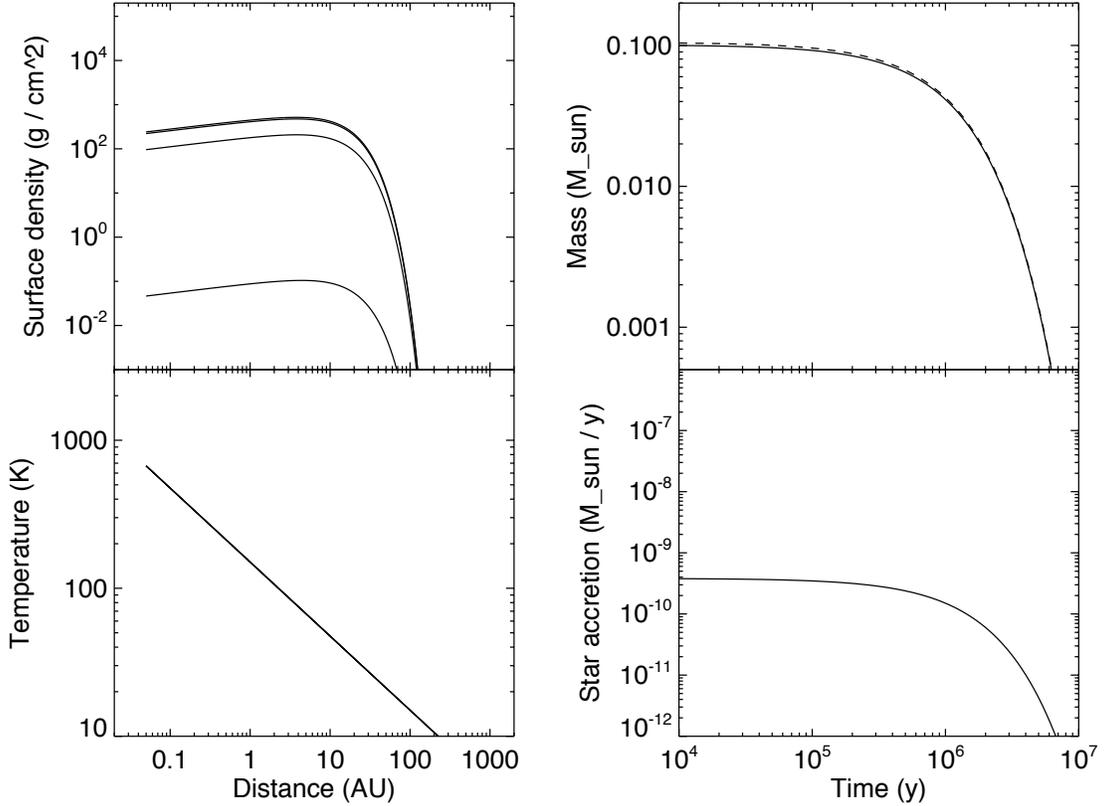}
\caption{Evolution of a disk with almost no viscosity and a slow disk wind ($v_0=10$ cm/s, $f_w=0.99$, and $K=1$). The initial mass and exponential radius are 0.1 solar masses and 15 AU, respectively. The upper, left panel shows the surface density at 0.01 My, 0.1 My, 1 My and 10 My, with lower curves corresponding to later times. The lower, left panel shows the temperature which is almost independent of time. The upper right panel shows the total disk mass (solid curve), and a rough estimate for the disk mass assuming it is entirely radiative (dashed curve). The lower right panel shows the mass accretion rate onto the star.}
\end{figure}

These differences are due to the mass ejected in the disk wind. Unlike the previous cases, the inward mass flux is no longer approximately independent of radius. Instead, mass is continually being removed by the wind as material flows inwards. Thus, the mass flux declines substantially as one moves closer to the star. This depresses the surface density in the inner disk compared to the previous cases, making the profile flatter and progressively reducing $\Sigma$ as $r$ decreases. 

Because the disk wind removes much of the inflowing mass, the mass accretion rate onto the star is much lower than in Figures~1 and 2. The stellar mass accretion rate (assuming the inner edge of the disk is at 0.05 AU) is 20--30 times smaller in Figure~3 at early times. The reduction is greater still at later times, with the accretion rate falling below $10^{-11} M_\odot$/y after a few My. The very low mass accretion rate late in the evolution is due to the combination of mass ejected by the disk wind, and the efficient removal of mass at early times due to the low degree of radial spreading. 

The temperature profile in Figure~3 is also notably different than the previous cases. Temperatures in the inner disk are substantially lower, and never reach the dust evaporation temperature. The temperature is no more than about a factor of 2 higher than that due solely to stellar irradiation. The nearly flat surface density profile also means that the temperature profile in the viscously heated region has almost the same slope as in the region heated by stellar irradiation. Temperatures in the inner disk are lower than the previous example due to the much lower surface density, which reduces the amount of viscous heating, and also reduces the vertical optical depth of the disk.

The cases shown in Figures 2 and 3 are similar in a few ways. The degree of viscous spreading is similar, since this is controlled by $f_w$ rather than $K$. The disk mass evolution is also broadly similar since the inflow velocities and radial spreading are similar in both cases.

%
%
\subsection{Nearly Laminar Disk with a Slow Wind}
The final case we consider is a disk with almost no viscosity and a slow disk wind. The initial mass and exponential radius are 0.1 solar masses and 15 AU, respectively, as in all the previous cases. However, the inflow velocity at 1 AU and 150 K is reduced so that $v_0=10$ cm/s. We set $f_w=0.99$ so that almost all the inflow is caused by the disk wind. (In this case, the viscosity is equivalent to $\alpha\simeq 3\times 10^{-5}$ for an alpha disk.) We also set $K=1$, the same as the previous example, so that a substantial amount of mass is ejected by the disk wind. The initial surface density at 1 AU in this case is $\Sigma_0=461$ g/cm$^2$.

The upper, left panel of Figure 4 shows the surface density evolution of the disk in this case. In this example, the rate of mass loss due to the wind is sufficient to reverse the slope of the surface density in the inner disk. Unlike the previous cases, $\Sigma$ {\em increases\/} with distance out to about 10 AU throughout the disk lifetime. There is essentially no radial spreading of the disk due to the very low level of viscosity. 

Early in the evolution, the surface density declines more slowly over time than in Figure~3 due to the smaller value of $v_0$. However, the lack of radial spreading means that the initial rate of mass loss is maintained for longer, so the surface density declines more rapidly at later times compared to Figure~3. This can also be seen in the disk mass evolution, shown in the upper, right panel of the figure. At early times, the mass in Figure~4 declines more slowly than in Figure~3, but the situation is reversed later on. After 10 million years, the total disk mass is reduced to $2.7\times 10^{-5}M_\odot$ in Figure~4 compared with $1.4\times 10^{-4} M_\odot$ in Figure~3.

The very low degree of viscosity in this example means that viscous heating is negligible. As a result, the temperature is essentially controlled by stellar irradiation at all radii. The temperature profile is static, as can be seen in the lower, left panel of the figure, since the stellar luminosity is assumed to be constant.

The mass accretion rates onto the star, shown in the lower, right panel, are the lowest of all the cases we have considered. The stellar accretion rate is always less than $5\times 10^{-10}M_\odot$/y, and becomes even smaller after 1 My. The positive surface density slope in the inner disk means that the surface density close to the star is quite small. This, combined with the lower value of $v_0$, ensures that the stellar mass accretion rate is very low. 

We note that this example was chosen to demonstrate that a disk wind can lead to a surface density that increases with distance, as found by \citet{suzuki:2016} in some cases. However, the corresponding stellar accretion rates are lower than those observed for some pre-main sequence stars \citep{hartmann:2016}, so this extreme case may be unrealistic for those systems.

The cases shown in Figures 3 and 4  demonstrate a feature of protoplanetary disks that may be important for interpreting observations of real disks. For purely viscous disks, there is a correlation between the disk mass, radius, and stellar accretion rate that can be used to constrain the viscosity \citep{rafikov:2017}. This correlation is broken when a disk has a wind that removes a substantial fraction of the mass.

\begin{figure}
\plotone{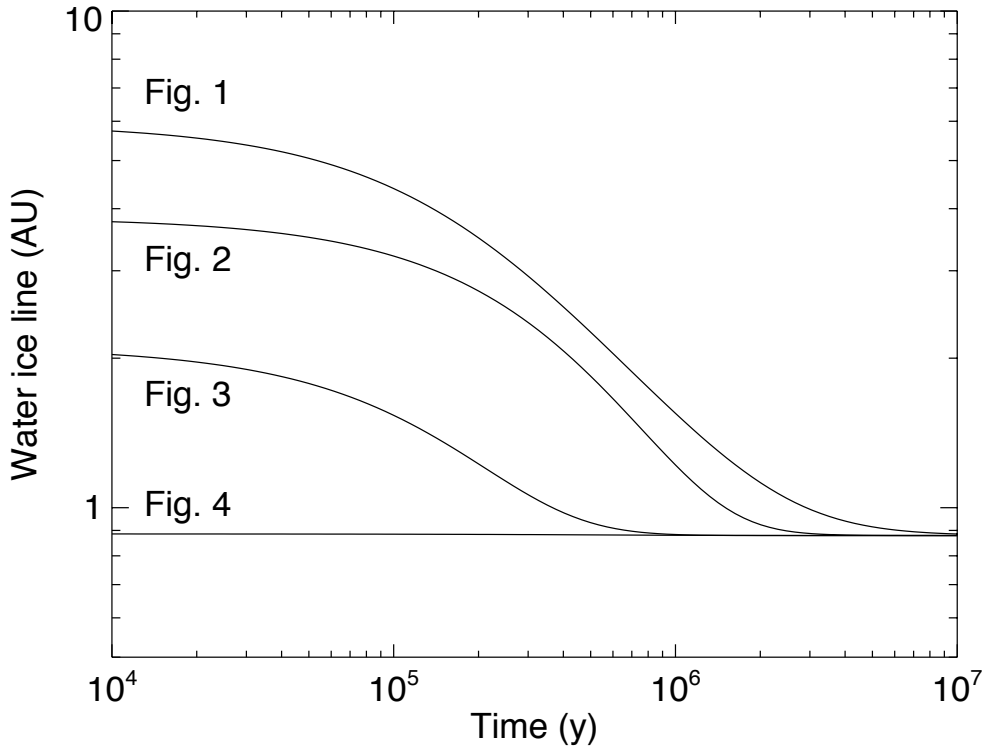}
\caption{Location of the water ice condensation distance (midplane temperature of 160~K) as a function of time for the cases shown in Figures~1--4.}
\end{figure}

%
%
\subsection{Location of the Ice Line}
The presence of a disk wind is likely to have many implications for planet formation, some of which have been explored elsewhere \citep{suzuki:2016, hasegawa:2017, ogihara:2018}. A detailed analysis is beyond the scope of this paper, but we briefly discuss one aspect here as an example of the capability of the model described above.

The examples discussed in the previous sections show that the disk temperature profile and its variation over time can depend on the presence and strength of a disk wind. This in turn affects the location of condensation fronts within the disk, with implications for the chemical composition of solids at a given location.  With this in mind, we examine the location of the most important condensation front, the water ice line. For simplicity, we ignore pressure dependence, and assume the ice line is located where the midplane temperature is 160~K.

Figure~5 shows the radial position of the ice line versus time for the models shown in Figures~1--4. The overall trend in each case is the same: the ice line moves inwards over time as the disk loses mass and cools. However, the motion of the ice line varies a good deal between the different cases. In the purely viscous disk shown in Figure~1, the ice line begins near 6 AU from the star, and its location changes substantially over the next 2 My. In this as in all of the cases, the ice line ends up just inside 1 AU, but it doesn't move inside 1 AU until after 3 My. 

In the case with the fast disk wind, shown in Figure 2, the degree of viscous heating is less than in Figure~1. The ice line begins at about 4 AU, and moves inwards somewhat more slowly than the purely viscous case. The ice line crosses 1 AU after about 1.5 My, ending up at about 0.9 AU as before. In the slow-disk-wind case shown in Figure~3, the low surface densities in the inner disk reduce viscous heating still further. In this case, the ice line starts just outside 2 AU, and crosses 1~AU at about 0.4 My, changing very little after that. Finally, the nearly laminar disk shown in Figure~4 has almost no viscous heating, and the ice line is essentially static at 0.9 AU throughout the evolution.

%
%
\section{Summary}
In this paper, we have derived an analytic model for an evolving protoplanetary disk driven by a combination of viscosity and  a disk wind. The disk is heated by irradiation from the central star and energy generated by viscosity. The model gives the midplane temperature and surface density as a function of time and distance from the star.

The model formulae are summarized in Section 3. The main features of the model are
\begin{enumerate}
\item The disk evolution is controlled by 3 factors: (i) the inflow velocity at a reference distance and temperature, (ii) the fraction of this inflow that is caused by the disk wind, and (iii) the rate of mass ejection by the disk wind compared to the inflow through the disk.

\item In a purely viscous disk, the surface density in the inner disk declines over time due to radial spreading and mass accretion onto the star. Radial spreading is more important than mass accretion in this regard, and the disk can still have a substantial mass at late times.

\item In viscous disks, temperatures in the inner disk are mainly caused by viscous heating, while stellar irradiation dominates in the outer disk. The boundary between these regions moves inwards over time. Close to the star, the temperature profile becomes nearly flat as dust evaporates and the opacity falls.

\item The presence of a strong wind changes many of these features. Radial spreading is reduced, and more of the mass remains in the inner disk, leading to more efficient accretion onto the star. The total disk mass at late times is much lower than in the purely viscous case. Temperatures in the inner disk are lower because viscous heating is relatively less important.

\item If the wind ejects a substantial amount of mass, the surface density profile becomes shallower. In extreme cases, the surface density can {\em increase\/} with distance in the inner disk. Surface densities in the inner disk are small compared to disks with negligible mass ejection.

\item The low surface density, combined with weak viscous heating, means that stellar irradiation is the main factor determining temperatures for most of the evolution in this case. As a result, the radial variation of the ice line is small compared to purely viscous disks.
\end{enumerate}

\acknowledgments
I thank an anonymous reviewer for helpful comments that improved this paper.

\begin{figure}
\plotone{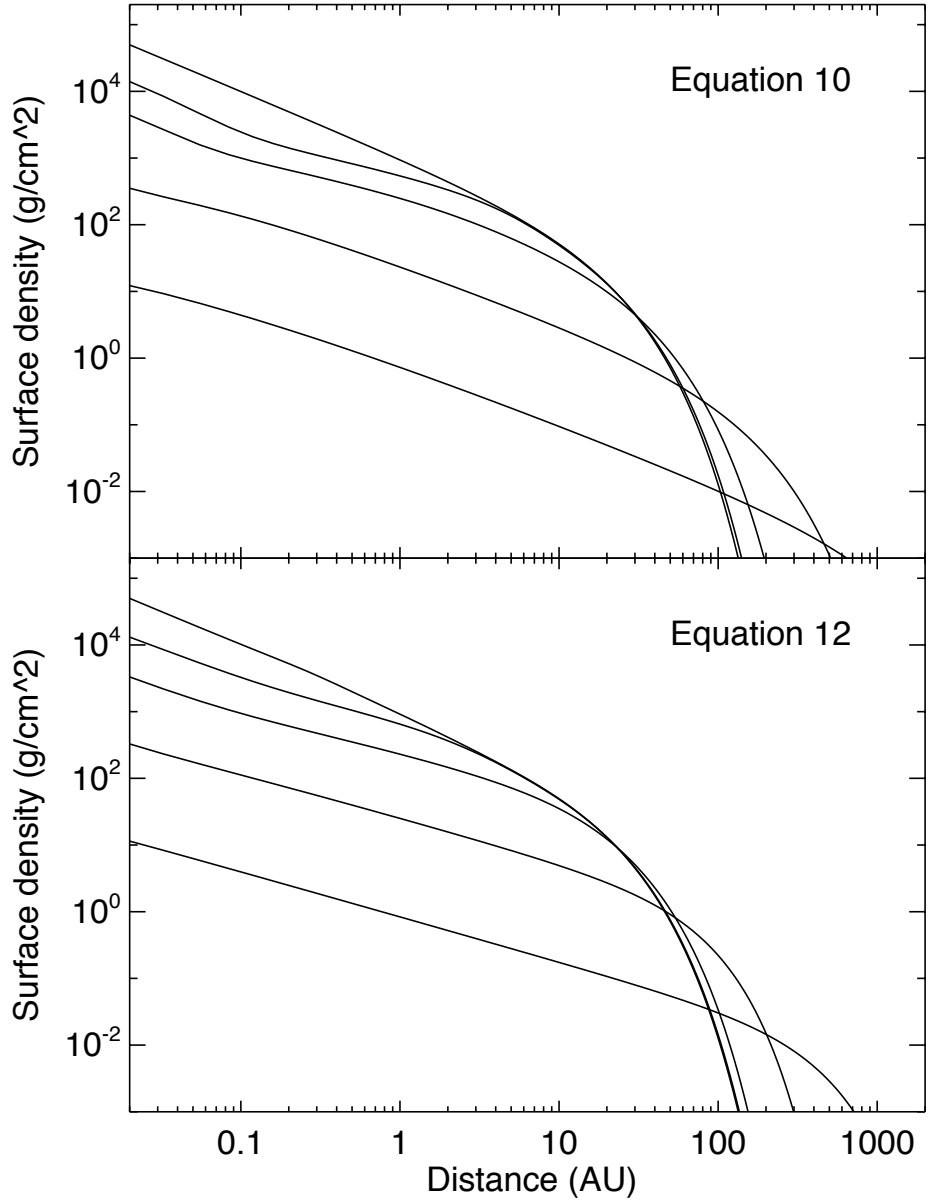}
\caption{The surface density of a disk with $v_0=30$ cm/s, $f_w=0.2$ and $K=0.4$ calculated numerically according to the true evolution, given by Eqn.~\ref{eq_true} (upper panel), and the approximation used in the analytic model, given by Eqn~\ref{eq_simplified} (lower panel). The curves correspond to the initial state, 0.01 My, 0.1 My, 1 My and 10 My, with lower curves corresponding to later times.}
\end{figure}

%
%
\appendix
In this appendix, we examine the effect of the approximations made to Eqn.~\ref{eq_true} in Section 2 that allow the disk evolution equations to be solved analytically.

Figure 6 shows numerical integrations of the original equation for the surface density evolution (Eqn.~\ref{eq_true}) and the modified expression (Eqn.~\ref{eq_simplified}) that was used to derive the analytic model used in the rest of the paper. In this case, we consider a disk that where the evolution is mainly driven viscosity. A modest disk wind is present with some mass loss associated with this wind. The model parameters are $v_0=30$ cm/s, $f_w=0.2$ and $K=0.4$.

The upper panel of Figure 6 shows the evolution according to Eqn.~\ref{eq_true}, calculated numerically using a radial grid with 1000 cells, with cell width proportional to $r^{1/2}$, and inner and outer boundaries at 0.005 and 3000 AU. The 5 curves show the initial surface density profile, and the profile at $10^4$, $10^5$, $10^6$ and $10^7$ years, with lower curves corresponding to later times. The lower panel of the figure shows the evolution according to Eqn.~\ref{eq_simplified} using the same numerical set up.

A few differences are apparent between the two cases. For example, the slope of the surface density profile varies somewhat more with radius in the inner disk in the upper panel  compared to the lower panel. The radial spreading of the outer edge is also somewhat different in the two cases. On the whole, however, the evolution is similar both qualitatively and quantitatively.

\begin{figure}
\plotone{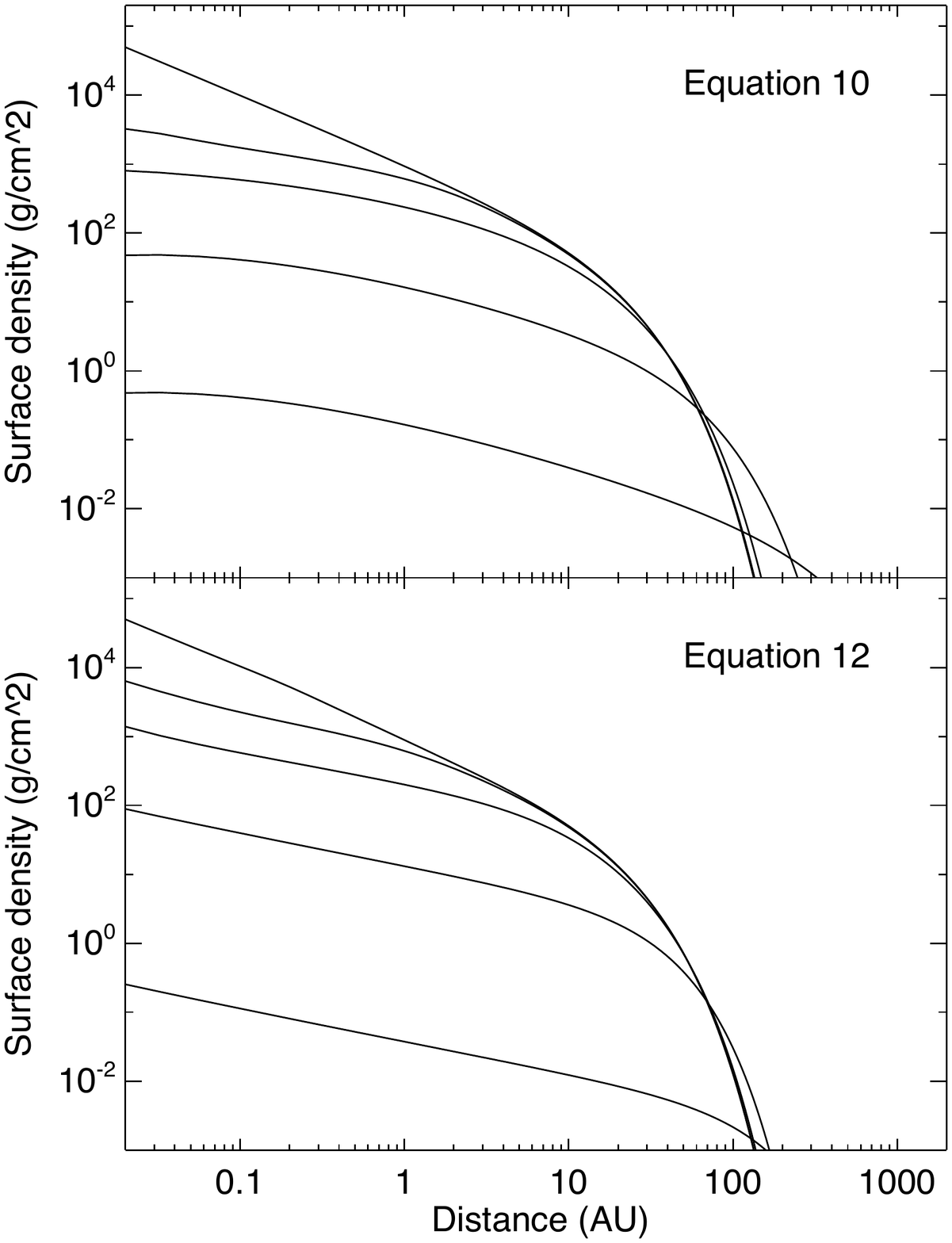}
\caption{The surface density of a disk with $v_0=30$ cm/s, $f_w=0.8$ and $K=0.4$ calculated numerically according to the true evolution, given by Eqn.~\ref{eq_true} (upper panel), and the approximation used in the analytic model, given by Eqn~\ref{eq_simplified} (lower panel).The curves correspond to the initial state, 0.01 My, 0.1 My, 1 My and 10 My, with lower curves corresponding to later times.}
\end{figure}

Next, we consider a case dominated by a disk wind with the same degree of mass loss as the previous example. The model parameters as $v_0=30$ cm/s, $f_w=0.8$ and $K=0.4$. The results according to Eqns.~\ref{eq_true} and \ref{eq_simplified} are shown in the upper and lower panels of Figure~7 respectively.

The differences between the two panels of Figure~7 are a little more apparent than the previous case, but still modest. The modified model (Eqn.~\ref{eq_simplified}) tends to underestimate the spreading of the disk's outer edge. It also leads to a steeper slope of the surface density profile inside about 0.3 AU, and overestimates the mass loss at late times somewhat. As in Figure~6 however, the differences are minor.

Finally we examine a disk dominated by a wind with a large amount of mass loss driven by the wind. In this case the model parameters are $v_0=30$ cm/s, $f_w=0.8$ and $K=0.8$, and the results, calculated numerically, are shown in Figure~8.

The trends shown in the previous two cases continue here, but the effects are more pronounced. Whereas Eqn.~\ref{eq_simplified} generates a disk in which the surface density always decreases with distance from the star, Eqn.~\ref{eq_true} leads to a modest increase in surface density with distance in the inner disk. These differences are largely confined to the region inside 1 AU, and the behavior at larger distances is similar in the two cases.

As before, the degree of radial spreading is somewhat larger when using Eqn.~\ref{eq_simplified}, and this equation tends to yield surface densities that are too low at 10 My compared to Eqn.~\ref{eq_true}.

We conclude that the modified equation used to derive the analytic disk model provides a good approximation to the true evolution in disks dominated by viscosity or a disk wind provided that the degree of mass loss associated with the wind is not too large.

When the disk is dominated by a disk wind and the associated mass loss is large, the results associated with the analytic model should be treated with caution in three respects. Firstly, the slope of the surface density profile in the inner disk (inside about 1 AU) is likely to be more positive than predicted. Secondly, the spreading of the outer edge of the disk will probably be less than suggested by the model. The model may also underestimate the surface density at late times.

Despite these caveats, we note that current uncertainties in the physics of disk evolution are almost certainly larger than the differences between Eqns.~\ref{eq_true} and \ref{eq_simplified}. Thus, attempts to provide a better approximation to Eqn.~\ref{eq_true} are probably not worthwhile at present.

\begin{figure}
\plotone{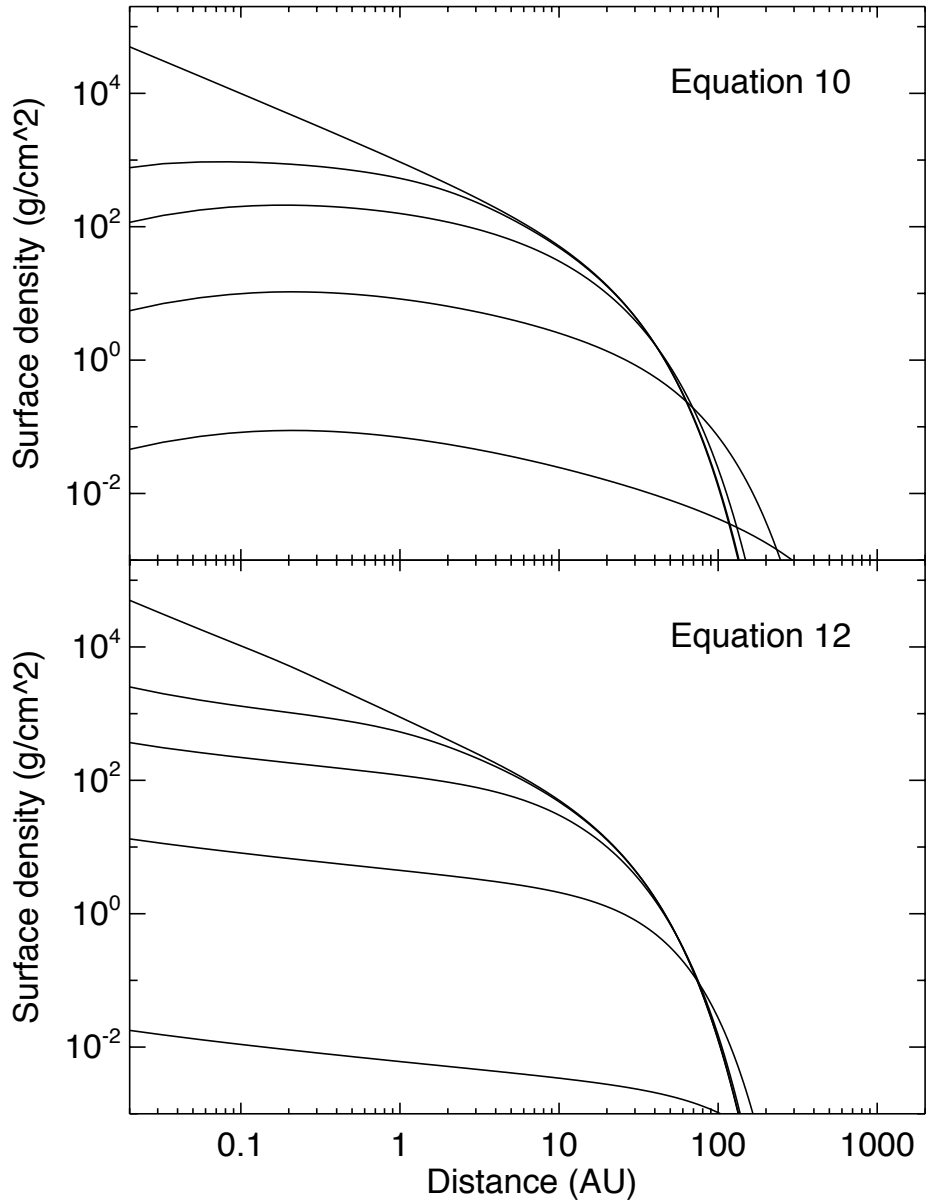}
\caption{The surface density of a disk with $v_0=30$ cm/s, $f_w=0.8$ and $K=0.8$ calculated numerically according to the true evolution, given by Eqn.~\ref{eq_true} (upper panel), and the approximation used in the analytic model, given by Eqn~\ref{eq_simplified} (lower panel). The curves correspond to the initial state, 0.01 My, 0.1 My, 1 My and 10 My, with lower curves corresponding to later times.}
\end{figure}

%
%

\end{document}